\documentclass{nature_plusfigure}

\usepackage{graphicx,float}
\usepackage{subcaption}
\usepackage{amsmath,amssymb,xcolor,url}
\usepackage{multirow}
\usepackage{longtable,tabularx,tablefootnote}
\usepackage{upgreek}
\usepackage{url}
\graphicspath{{./}{figures/}}
\renewcommand{\thefootnote}{\textit{\alph{footnote}}}

\usepackage[none]{hyphenat}
\usepackage{multibib}
\usepackage{xspace}
\newcites{New}{References}

\def \grb {\mbox{GRB\,231115A}}

\def\ltsima{$\; \buildrel < \over \sim \;$}
\def\lsim{\lower.5ex\hbox{\ltsima}}
\def\gtsima{$\; \buildrel > \over \sim \;$}
\def\gsim{\lower.5ex\hbox{\gtsima}}
\def\msun{M_{\odot}}

\newcommand{\flux}{erg\,cm$^{-2}$\,s$^{-1}$\xspace}
\newcommand{\lum}{erg\,s$^{-1}$\xspace}
\newcommand{\mrate}{$M_{\odot}$\,yr$^{-1}$\xspace}
\newcommand{\integral}{\textit{INTEGRAL}\xspace}
\newcommand{\xmm}{\textit{XMM-Newton}\xspace}
\newcommand{\swift}{\textit{Swift}\xspace}
\newcommand{\coord}[8]{R.A.\,=\,${#1}^{\rm h}$ ${#2}^{\rm m}$ ${#3}^{\rm s}{#4}$, Dec.\,=\,${#5}^\circ$ ${#6}'$ ${#7}''{#8}$\xspace}

\newcommand{\mn}{{Mon. Not. R. Astron. Soc.}}

\newcommand{\mnras}{\mn}

\newcommand{\apj}{{Astrophys. J.}}
\newcommand{\apjl}{{Astrophys. J. Lett.}}
\newcommand{\apjs}{{Astrophys. J. Supp.}}

\newcommand{\aap}{{Astron. Astrophys.}}
\newcommand{\aapr}{{Astron. Astrophys. Rev.}}
\newcommand{\nat}{{Nature}}

\newcommand{\pasp}{{Pub. Ast. Soc. Pac.}}

\newcommand{\araa}{Annual Review of Astronomy and Astrophysics}

\usepackage[symbol]{footmisc}
\renewcommand{\thefootnote}{\fnsymbol{footnote}}

\usepackage{cleveref}

\usepackage[none]{hyphenat}
\usepackage{multibib}
\newcites{New}{References}
\usepackage{subcaption} 

\title{A magnetar giant flare in the nearby starburst galaxy M82}

\author{Sandro Mereghetti$^{1,*}$, 
 Michela Rigoselli$^{1}$, 
 Ruben Salvaterra$^{1}$, 
 Dominik Patryk Pacholski$^{1,2}$, 
 James Craig Rodi$^{3}$, 
 Diego Gotz$^{4}$, 
 Edoardo Arrigoni$^{5,1}$, 
 Paolo D'Avanzo$^{6}$, 
 Christophe Adami$^{7}$, 
 Angela Bazzano$^{3}$,  
 Enrico Bozzo$^{8,6}$,  
 Riccardo Brivio$^{9,6}$, 
 Sergio Campana$^{6}$, 
 Enrico Cappellaro$^{10}$, 
 Jerome Chenevez$^{11}$, 
 Fiore De Luise$^{12}$, 
 Lorenzo Ducci$^{13,8}$, 
 Paolo Esposito$^{14,1}$, 
 Carlo Ferrigno$^{8,6}$, 
 Matteo Ferro$^{9,6}$, 
 Gian Luca Israel$^{15}$, 
 Emeric Le Floc’h$^{4}$, 
 Antonio Martin-Carrillo$^{16}$, 
 Francesca Onori$^{12}$, 
 Nanda Rea$^{17,18}$, 
 Andrea Reguitti$^{6,10}$, 
 Volodymyr Savchenko$^{19}$, 
  Damya Souami$^{20}$,
 Leonardo Tartaglia$^{12}$,  
 William Thuillot$^{21}$, 
 Andrea Tiengo$^{14,1}$, 
 Lina Tomasella$^{10}$, 
 Martin Topinka$^{22}$, 
 Damien Turpin$^{4}$, 
 Pietro Ubertini$^{3}$  
 }

\begin{document}

\maketitle

\begin{small}
\begin{affiliations}
\label{sec:affiliations}

\item INAF - Istituto di Astrofisica Spaziale e Fisica Cosmica di Milano, Via A. Corti 12, 20133 Milano, Italy

\item Universit\`a degli Studi di Milano Bicocca, Dipartimento di Fisica G. Occhialini, Piazza della Scienza 3, 20126 Milano, Italy

\item INAF - Istituto di Astrofisica e Planetologia Spaziali di Roma, Via del Fosso del Cavaliere 100, 00133 Roma, Italy  

\item Université Paris-Saclay, Université Paris Cité, CEA, CNRS, AIM, 91191, Gif-sur-Yvette, France

\item Universit\`a degli Studi di Milano, Dipartimento di Fisica, Via Celoria 16, 20133 Milano, Italy

\item INAF - Osservatorio Astronomico di Brera, Via E. Bianchi 46, 23807 Merate, Italy

\item Aix Marseille Univ, CNRS, CNES, LAM, Marseille, France

\item University of Geneva, Department of Astronomy, Chemin d'Ecogia 16, Versoix 1290, Switzerland

\item Universit\`a dell’Insubria, Dipartimento di Scienza e Alta Tecnologia, Via Valleggio 11, 22100 Como, Italy

\item INAF - Osservatorio Astronomico di Padova, Vicolo dell’Osservatorio 5, 35122 Padova, Italy 

\item DTU Space, Technical University of Denmark, Elektrovej 327, 2800 Kongens Lyngby, Denmark

\item INAF - Osservatorio Astronomico d'Abruzzo, Via M. Maggini snc, 64100 Teramo, Italy

\item Institut fuer Astronomie und Astrophysik Tuebingen, Sand 1, Tuebingen, Germany

\item Scuola Universitaria Superiore IUSS Pavia, Piazza della Vittoria 15, 27100 Pavia, Italy

\item INAF - Osservatorio Astronomico di Roma, 00078 Monte Porzio Catone, Italy

\item School of Physics and Centre for Space Research, University College Dublin, Belfield, Dublin 4, Dublin, Ireland

\item Institute of Space Sciences (CSIC-ICE), Campus UAB,
Carrer de Can Magrans s/n, 08193 Barcelona, Spain

\item Institut d'Estudis Espacials de Catalunya (IEEC), Carrer Gran Capit\`a 2--4, 08034 Barcelona, Spain

\item École polytechnique fédérale de Lausanne, Lausanne, Switzerland.  

\item{LESIA, Observatoire de Paris, Université PSL, CNRS, Sorbonne Université, Université de Paris, 5 place Jules Janssen, F-92195 Meudon, France}

\item Institut de mecanique celeste et de calcul des ephemerides (IMCCE)
UMR 8028 du CNRS - Observatoire de Paris 77 Av. Denfert Rochereau 75014 Paris, France

\item INAF - Osservatorio Astronomico di Cagliari, Via della Scienza 5, 09047 Selargius (CA), Italy\\


$^{*}$ Corresponding author: sandro.mereghetti@inaf.it

\end{affiliations}
\end{small}

\renewcommand{\thefootnote}{\textit{\alph{footnote}}}
\newcommand\brobor{\smash[b]{\raisebox{0.6\height}{\scalebox{0.5}{\tiny(}}{\mkern-1.5mu\scriptstyle-\mkern-1.5mu}\raisebox{0.6\height}{\scalebox{0.5}{\tiny)}}}}
\captionsetup[table]{name=Table}

\begin{abstract}

Giant flares, short explosive events releasing up to 10$^{47}$ erg of energy   in the gamma-ray band in less than one second, are the most spectacular manifestation of magnetars,  young neutron stars powered by a very strong magnetic field, 10$^{14-15}$ G in the magnetosphere and possibly higher in the star interior\cite{Mereghetti2008,Kaspi2017}.
The rate of occurrence of these rare flares is poorly constrained, as only three  have been seen from three different magnetars in the Milky Way\cite{Hurley1999,Palmer2005} and in the Large Magellanic Cloud\cite{Mazets1979} in $\sim$50 years since the beginning of gamma-ray astronomy. This sample can be enlarged by the discovery of extragalactic events, since for a fraction of a second giant flares reach peak luminosities above 10$^{46}$ \lum, which makes them visible by current instruments up to a few tens of Mpc.  However, at these distances they appear similar to, and difficult to distinguish from, regular short gamma-ray bursts (GRBs). The latter are much more energetic events, 10$^{50-53}$ erg,  produced by compact binary mergers and originating at much larger distances\cite{Davanzo2015}. Indeed, only a few short GRBs have been proposed\cite{Frederiks2007b,Mazets2008,Svinkin2021,Burns2021,Trigg2023},  with different levels of confidence, as magnetar giant flare candidates in nearby galaxies. 
Here we report the discovery 
of a short GRB positionally coincident with the central region of the starburst galaxy M82\cite{Forster2003}.
Its spectral and timing properties, together with the limits on its X-ray and optical counterparts obtained a few hours after the event and the lack of an associated gravitational wave signal,  qualify with high confidence this event as a giant flare from a magnetar in M82. 

\end{abstract}

\begin{figure}[ht]
 \centering
  \includegraphics[width=18cm]
{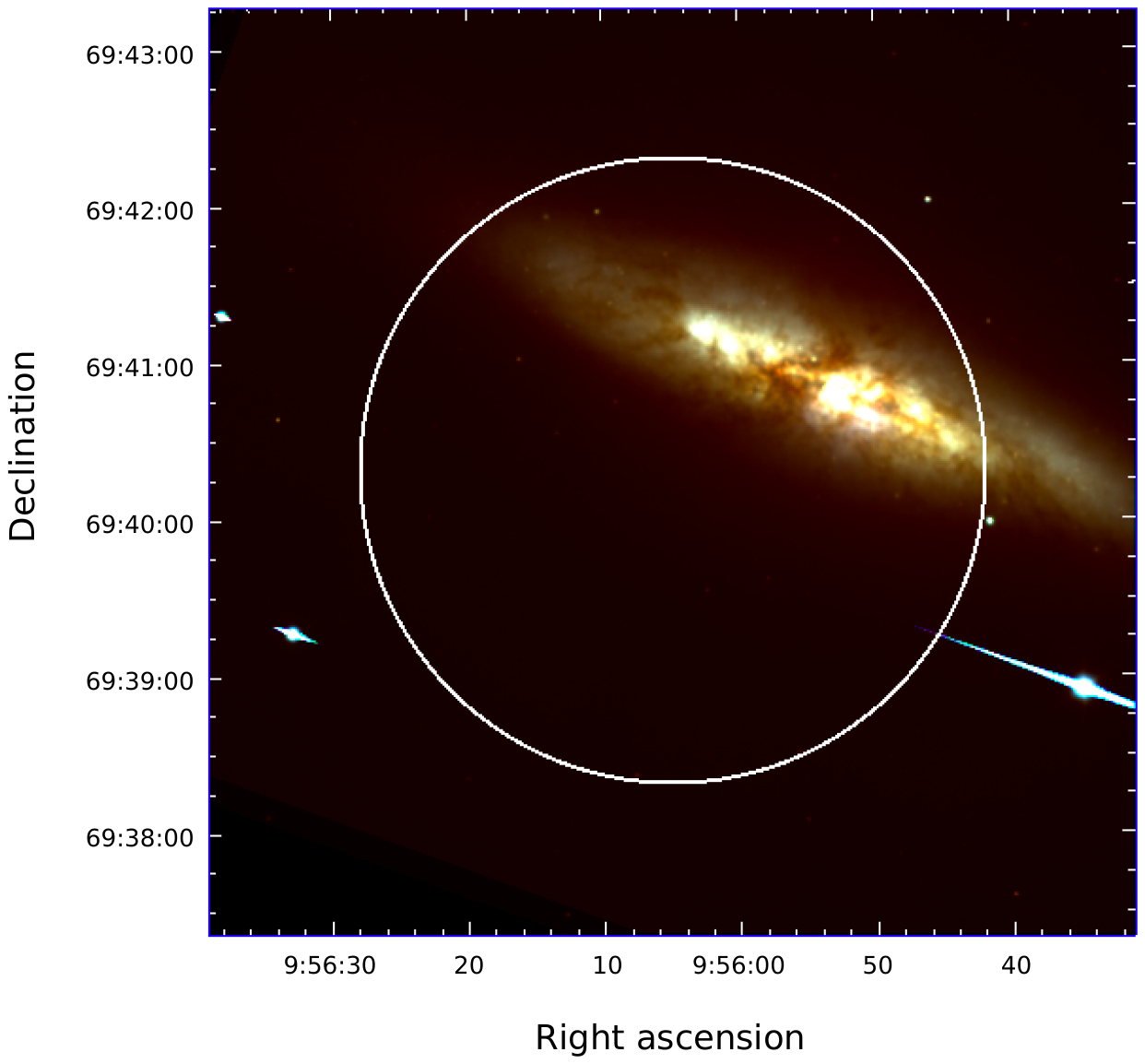}
   \caption{\textbf{Optical image of M82} (from data obtained with the TNG). RGB scale: z-band (red), i-band (green), r-band (blue). The 90\% c.l.\ error circle of \grb\ has a radius of 2 arcmin. }
  \label{fig:imaopt}
\end{figure}

Instruments on board the \integral satellite detected  \grb\ on 2023 November 15 at 15:36:20.7 UT\cite{Mereghetti2023}.
This short burst (duration $T_{90}$ = 93 ms)  was seen also by other satellites\cite{Dalessi2023,Cheung2023,Xue2023,Frederiks2023,Wang2023}, but only \integral could promptly associate it to M82 thanks to a localization at few arcmin level, publicly distributed by the \integral Burst Alert System (IBAS\cite{Mereghetti2003}) only 13 s after the burst detection. 
The burst occurred in the field of view of the IBIS coded mask instrument\cite{Ubertini2003}, at coordinates
R.A.=149.0205 deg, Dec.=+69.6719 deg (J2000, 2 arcmin 90\% c.l.\ radius). This position is consistent with, and supersedes,  the one derived using the preliminary satellite attitude available in near real time\cite{Mereghetti2023}. 
The position of \grb\ coincides with the nearby starburst galaxy M82\cite{Forster2003} (see Figure \ref{fig:imaopt}). Notably, the central region of the galaxy, where most star formation activity occurred, is inside the \integral/IBIS error region. 
Considering that the total angular size of all the galaxies with apparent luminosity brighter than M82 (excluding the Magellanic Clouds and M31)
is $\sim$6000 arcmin$^2$, the chance alignment with \grb\ is 4$\times10^{-5}$. 
A more elaborate analysis\cite{Burns2023} indicates that the Bayes factor favouring a giant flare in M82 with respect to  a chance alignment of a short GRB in the background is a factor 30 larger than that of the previous best candidate  (GRB 200415A possibly associated to NGC 253\cite{Svinkin2021}), making \grb\ by far the most compelling case for a magnetar giant flare outside the local group of galaxies.

\begin{figure}[ht]
 \centering
 \includegraphics[width=16cm]{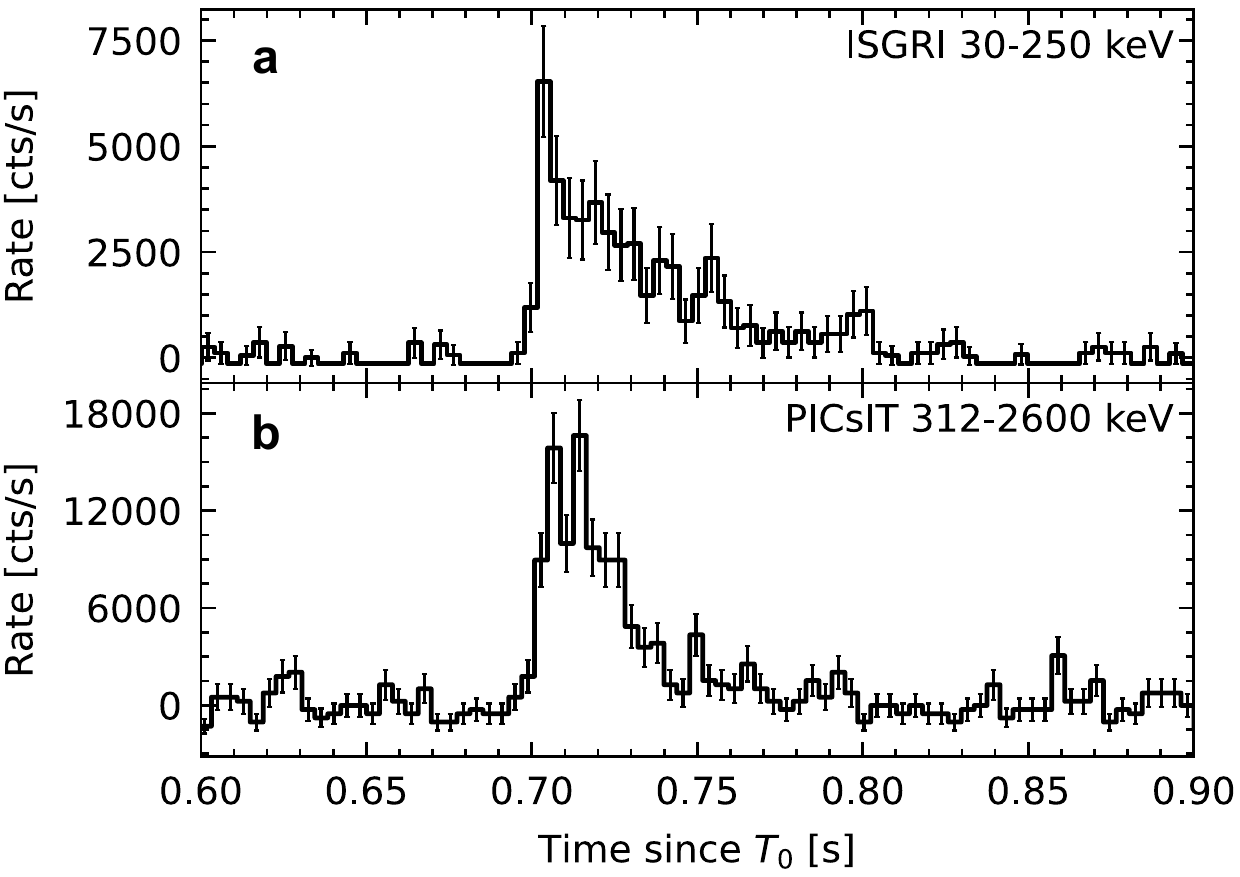}
  \vspace{-5cm}
    \caption{ \textbf{Light curves of \grb.} Background-subtracted light curves (errors are at 1$\sigma$) obtained with the ISGRI detector in the $30-250$ keV energy range (\textbf{a}) and with the PICsIT detector in the $312-2600$ keV energy range (\textbf{b}). Time is referred to T$_0$ = 2023-11-15 15:36:20 UT (time at the \integral position; the burst reached instruments on low Earth orbit satellites about 0.4 s later).   } 
    \label{fig:lcurve}
\end{figure}

The light curves plotted in Figure \ref{fig:lcurve} show that
\grb\ was clearly visible in the two IBIS detectors, which operate in different energy ranges:  ISGRI\cite{Lebrun2003} (20 keV -- 1 MeV) and PICsIT\cite{Labanti2003} (175 keV -- 15 MeV).  
The ISGRI and PICsIT spectra of the burst, extracted from the time interval with a significant detection in both detectors (from T$_0$+0.687 s to T$_0$+0.748 s, with  T$_0$=15:36:20.0 UT), have been jointly fit and are well described by an exponentially cut-off power law
with photon index $\alpha=0.04^{+0.27}_{-0.24}$, 
peak energy $E_{p} = 551^{+81}_{-59}$ keV,
and flux $F_{30-2600}=(7.2^{+0.6}_{-0.7})\times10^{-6}$
\flux in the  30 keV -- 2.6 MeV range 
(see Methods sect.~\ref{sec:integral}).
Assuming this spectral shape for the whole duration of the burst, the fluence in the same energy range is  $(6.3\pm0.5)\times10^{-7}$  erg cm$^{-2}$.

M82 was in the field of view of IBIS starting from 9.2 hours before up to 11.8 minutes after the burst, as well as during a target of opportunity (ToO) observation performed in the following three days, but no other bursting or persistent emission besides \grb\ was detected.
Integrating all the data of the ToO observation, we obtained an upper limit of  $2.6\times10^{-11}$ \flux on the 30--100 keV persistent emission at the position of \grb . 
Also, no other bursts from  M82 were detected   
in all the available ISGRI archival data obtained since 2003, which provided a total  exposure of more than 16 Ms
(see Methods sect.~\ref{sec:integral}).

At the distance of M82 (3.6 Mpc\cite{Freedman1994}), the \grb\ fluence measured with IBIS implies an emitted isotropic energy $E_{\rm iso}$ = $10^{45}$ erg (1 keV -- 10 MeV), well below the typical value for a short GRB, but 
consistent with the one of the   initial pulses of the three giant flares securely associated with magnetars\cite{Hurley1999,Palmer2005,Mazets1979}.  
The short duration and hard spectrum of \grb\ are also in agreement with the properties seen in the initial pulses of magnetar giant flares. 
In the three  giant flares securely associated with magnetars, the initial short and hard pulse was followed by a softer  tail, with a duration of a few minutes and a maximum luminosity of  $\sim\!10^{42}$ \lum, characterized by a periodic modulation induced by the neutron star rotation. 
A similar feature would be too faint  to  be visible by IBIS at the distance of M82. This is shown in Figure~\ref{fig:sgr1806}, where we compare the limits obtained for \grb\ with the light curve of the most energetic Galactic giant flare ever observed  (the one emitted by SGR 1806--20 in 2004\cite{Palmer2005}) rescaled to a distance of 3.6 Mpc.

The field of \grb\ was promptly observed with the Neil Gehrels Swift Observatory starting 9 ks after the burst\cite{Osborne2023} and
a deeper X-ray observation  was carried out with the \xmm satellite, starting 0.7 days after the burst\cite{Rigoselli2023} (see Methods sect.~\ref{sec:xmm}). 
The comparison of the \swift and \xmm images with those obtained in  previous observations of M82 does not show evidence for new X-ray sources inside the \grb\ error region.  Due to the presence of unresolved X-ray emission from  the central part of the galaxy, the corresponding upper limit is position-dependent. 
The limit of $4\times10^{-14}$ \flux derived from the \xmm data, which applies to 60\% of the error region, is inconsistent with the flux at T$_0$+1 day of the large majority of X-ray afterglows of short GRBs detected with the \swift/XRT instrument (see Methods Figure~\ref{fig:afterglows}). Although we can not exclude a rapid afterglow drop, as observed in the class of so-called short lived short GRBs\cite{Sakamoto2009}, this X-ray upper limit disfavors the GRB interpretation. It is instead consistent with the quiescent X-ray emission of a magnetar in M82. In fact the  X-ray luminosity of known magnetars, also considering the peak values reached during outbursts\cite{CotiZelati2018}, never exceeds 10$^{36}$ \lum, corresponding to $6\times10^{-16}$ \flux at the  M82 distance.
  
Holding true the association of \grb\ with M82 on the basis of the remarkable spatial coincidence, the low value of $E_{\rm iso}$ could be reconciled with a short GRB origin only if the jet axis is pointed away from our direction, similar to the  case of the gravitational wave event GRB 170817A\cite{Abbott2017}, which produced the bright kilonova AT2017gfo in the optical/NIR band.
However, the deep upper limits of m\,$>$\,20.0 -- 24.0 mag (depending on the assumed position  in the \integral error circle) on the optical counterpart that we obtained starting at T$_0$+0.2 days (see Methods sect.~\ref{sec:optical}) exclude this possibility, both for an AT2017gfo-like event and for even fainter kilonovae (see Methods sect.~\ref{sec:optical}). 
We finally note that the binary merger of two compact objects at 3.6 Mpc would have produced a strong  signal in gravitational waves, at variance with the non-detection reported by the LIGO/Virgo/KAGRA Collaboration\cite{Ligo2023}.

The discovery of a young, active magnetar in M82, a starburst galaxy  characterized by a high star formation rate\cite{Leroy2019} is consistent with the origin of magnetars in core collapse supernova explosions\cite{Duncan1992}. 
The volumetric rate of giant flares with $E_{\rm iso}>4\times10^{44}$ erg  has been recently estimated\cite{Burns2021} as   $(3.8^{+4.0}_{-3.1})\times10^5$ Gpc$^{-3}$\,yr$^{-1}$. 
Assuming a direct link with the young stellar population of core collapse progenitors and considering  the total star formation rate of 4000 \mrate within 50 Mpc, this corresponds   to  
a rate of $R(E_{iso}>4\!\times10^{44}) = 0.05^{+0.05}_{-0.04}$ yr$^{-1}$ (\mrate)$^{-1}$. 
Given the star formation rate  of 7.1 \mrate in M82\cite{Leroy2019} and  assuming a  power-law distribution of the giant flares energies  with slope 1.7, 
the expected rate of giant flares with 
$E_{\rm iso}>10^{45}$ erg in M82 is $R_{\rm M82}(E_{iso}>10^{45})
= 0.19^{+0.19}_{-0.15}$ yr$^{-1}$.
Recent calculations based on relativistic hydrodynamical simulations\cite{Cehula2023} show that such ejecta are efficient sources for the nucleosynthesis of heavy elements through the r-process.  
Thus the giant flares with $E_{\rm iso}>10^{46}$ erg (each producing up to $10^{26}$ g of ejecta) can give  a yield of $\sim2\times10^{-9}$ $\msun$ yr$^{-1}$ of r-process elements in M82 through this channel.

GRB 051103 is another short GRB that has been proposed as a possible magnetar giant flare in the M81 group of galaxies\cite{Frederiks2007b}.  According to a detailed statistical analysis\cite{Burns2021}, M82 is the most likely host of GRB 051103, despite being slightly outside its 260 arcmin$^2$ localization region. 
The rate of giant flares derived above is fully consistent with the detection of two events from M82 in 20 years.
Therefore, starburst galaxies such as M82, which produce a significant population of young magnetars, appear as promising targets 
to constrain the energy distribution function of giant flares 
by means of long dedicated observations with future high sensitivity instruments.

\section*{References}


%

\begin{figure}[ht]
 \centering
 \includegraphics[width=18cm]{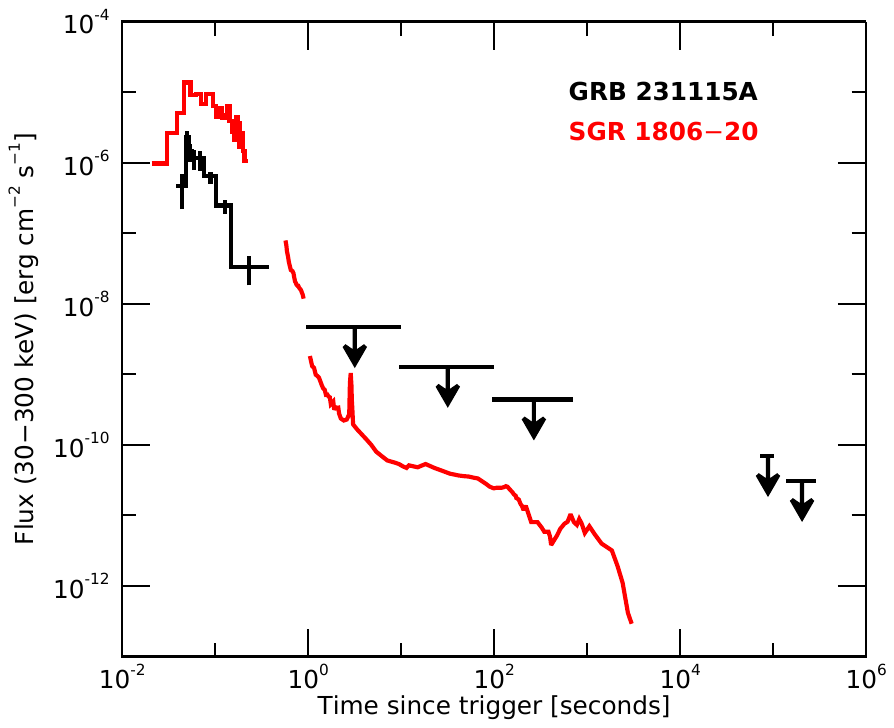}
 \vspace{-6cm}
    \caption{ \textbf{Comparison between \grb\  and the 2004 giant flare from SGR 1806--20}. Errors are at 1$\sigma$, upper limits are at 3$\sigma$.
    The light curve of SGR 1806--20 (red) has been rescaled to the distance of M82 (data from Helicon-Coronas-F\cite{Frederiks2007a}, Konus-Wind\cite{Frederiks2007a} and \integral SPI-ACS\cite{Mereghetti2005}). 
    Note that the 5.2 s periodicity in the   SGR 1806--20 light curve is not visible due to the used bin size. 
   }  \label{fig:sgr1806}
\end{figure}

\clearpage

\begin{methods}

\section{\integral data analysis and results}
\label{sec:integral}

IBIS is a $\gamma$-ray telescope based on two position-sensitive detection planes working in different overlapping energy ranges:  ISGRI\cite{Lebrun2003} (20 keV -- 1 MeV) and PICsIT\cite{Labanti2003} (175 keV -- 15 MeV).  
A  coded mask placed about 3.2 m above ISGRI provides imaging capabilities over a field of view of $\sim$30$\times$30 deg$^2$. 
\grb\ was detected at an off-axis angle of 4$^{\circ}$, inside the fully coded field of view. 
For the  data reduction and analysis, we  used version 11.2 of the Offline Science Analysis (OSA) software\cite{Goldwurm2003}, with the most recent calibration files, that properly account for the time evolution of the instrument response.

\subsection{Burst spectrum}
\label{sec:spectra}
We extracted an ISGRI spectrum, in 14 logarithmically spaced energy channels between $30-500$ keV, for the time interval from 15:36:20.687 to 15:36:20.748 UT.  
The PICsIT spectrum for the same time interval was extracted from the spectral-timing data mode in 8 predefined energy channels spanning $212-2600$ keV.
Data in spectral-timing mode (counts integrated over the whole detector in 3.9 ms time bins) do not provide imaging information (contrary to the ISGRI data). Therefore, the background count-rate in each PICsIT energy channel was calculated as the median count rate during the \integral pointing  (5013 s duration) containing the burst. A proper correction factor was applied to the PICsIT effective area to account for the angular distance between the IBIS pointing direction and   \grb.

We added 5\% systematic uncertainties to both spectra. The ISGRI and PICsIT spectra were fitted simultaneously using XSPEC\cite{Arnaud1996} version 12.12.1. 
A fit with a power law was unacceptable ($\chi^2=59.4$ for 17 degrees of freedom), 
while a power law with exponential cut-off,  defined as $F(E) = K E^{\alpha} \, \exp(-E(2+\alpha)/E_p)$, gave a good fit with 
photon index $\alpha = 0.04^{+0.27}_{-0.24}$,
peak energy $E_p  = 551^{+81}_{-59}$ keV, 
and $30-2600$ keV flux $F$ = $(7.2\pm0.6)\times10^{-6}$ \flux .
 
The ISGRI spectrum for the time interval corresponding to the whole GRB, from 15:36:20.687 UT to 15:36:20.833 UT, is less well constrained, but the best fit parameters are within the errors consistent with those of the ISGRI plus PICsIT spectrum. All the fit parameters are summarized in Table~\ref{tab:spectra}, where we give also the results obtained with a blackbody model.

\begin{table}
      \caption{{\bf Results of spectral fits of \grb.} Errors are at 1$\sigma$, fluxes are in the $30-2600$ keV energy range. Time intervals are in seconds after T$_0$ =  15:36:20 UT. }
    \centering
    \begin{tabular}{lcc}
    \hline
   Detectors      &  ISGRI+PICsIT &  ISGRI \\
   Time interval  & 0.687--0.748   &  0.687--0.833 \\
   \hline
       \multicolumn{3}{c}{Blackbody}\\   
       \hline
     $kT$ (keV)    &     $108^{+7}_{-5}$   & $68^{+12}_{-11}$\\
    Flux ($10^{-6}$ \flux) & $7.9^{+0.6}_{-0.7}$  & $2.1^{+0.6}_{-0.7}$\\
    $\chi^2$ / d.o.f. & $31.9/17$   & $16.4/12$ \\
    \hline
      \multicolumn{3}{c}{Cut-off power law}\\
   \hline
    $\alpha$    & $0.04^{+0.27}_{-0.24}$ &    $0.55^{+0.50}_{-0.42}$\\
    $E_p$ (keV) &  $551^{+81}_{-59}$ &    $355^{+158}_{-89}$ \\
    Flux ($10^{-6}$ \flux) & $7.2^{+0.6}_{-0.7}$  & $2.7\pm1.1$\\
    $\chi^2$ / d.o.f. & $17.2/16$ &   $14.9/11$\\
   \hline
   \end{tabular}   
    \label{tab:spectra}
\end{table}

\subsection{Follow up observation. }
  \label{sec:too}
  Immediately after the discovery of \grb\ we requested an \integral Target of Opportunity (ToO) observation that started only 21 hours after the burst and continued in the next satellite revolution.  The first part of the ToO  was done between 12:38 UT and 19:46 UT of November 16.  The next one from 07:21 UT of  November 17  to 23:09 UT of November 18, resulting in a total exposure of 162 ks divided in 57 pointings.
  Using the OSA 11.2 software, we produced the mosaics of the ISGRI images in the $30-100$ keV energy range for three periods: November 16, November 17-18, and November 16 to 18.
  No sources were detected at the position of \grb. The 3-$\sigma$ upper limits on the $30-100$ keV flux,  derived assuming a thermal bremsstrahlung spectrum with temperature $kT = 30$ keV, are
$6.5\times10^{-11}$ \flux for November 16, 
$2.9\times10^{-11}$ \flux for November 17-18,
and $2.6\times10^{-11}$ \flux for the sum of the two periods. 
We also derived an upper limit of $3.8\times10^{-10}$ \flux  on the flux in the time interval from T$_0$ + 1.7 s  to T$_0$ + 11.8 min,  during which the \grb\ position was in the IBIS field of view. All these upper limits are plotted in Figure~\ref{fig:sgr1806}, where it can be seen that, even a giant flare as energetic as the one emitted from SGR 1806--20, at the distance of M82 would have been undetectable by IBIS after the initial bright pulse.

\subsection{Search for past activity from M82.}
 \label{sec:search}

The region of M82 has been repeatedly observed with \integral, starting in November 2003.
After eliminating the time intervals  with strong and variable background (typically due to solar activity or to particles trapped in the Earth radiation belts when the satellite is close to perigee) we obtained about 16 millions of seconds of useful time with M82 in the IBIS field of view. 
We extracted ISGRI lightcurves in the $30-150$  keV energy range, using only detector pixels illuminated for more than 50\% from a source at the \grb\ position and searched for excess counts in the light curves over eight logarithmically spaced  timescales between 0.01 and 1.28 s (see \cite{Mereghetti2021} for more details).  All the burst candidates found in the light curves were then examined by making the sky images of the corresponding time intervals.  None of them showed the presence of a source at the M82 position. Through simulations we found that bursts down to a factor $\sim$5 fainter than \grb\ would have been detected in these data,  assuming the same time profile and spectrum.

\section{X-ray  observations}
\label{sec:xmm}

\swift pointed at M82, starting to collect  data 9.0 ks after the burst occurrence\cite{Osborne2023}. 
\swift/XRT observations were carried out in Photon Counting mode in the $9.0-39.2$ ks time interval, collecting 4.4 ks of data.
No new X-ray sources were detected within the \integral error region of \grb. 
The 3-$\sigma$ upper limit is position-dependent due to the  diffuse X-ray emission from the M82 galaxy.
Outside the galaxy, the upper limit on the count rate is $\sim 2-3\times 10^{-3}$ counts s$^{-1}$. 
Assuming a power-law spectrum with photon index $\Gamma=2$ and the Galactic column density of 
$6.5\times 10^{20}$ cm$^{-2}$ (the total Galactic absorption column in the line of sight of M82\cite{HI4PI})
this corresponds to a $0.3-10$ keV unabsorbed flux of $\sim 10^{-13}$ \flux.

A 47 ks long ToO observation of \grb\ was carried out with the \xmm satellite, starting on 2023 November 16 at 08:29:16 UT, about 16.9 hours after the burst  (Obs.ID 093239)\cite{Rigoselli2023}. 
The EPIC-pn\cite{Struder2001} and the two EPIC-MOS\cite{Turner2001} cameras were operated in Full window mode, with the thin optical-blocking filter. 
We processed the data with the \textsc{epproc} and \textsc{emproc} pipelines of version 18 of the Science Analysis System (\textsc{sas})\cite{Gabriel2004} and the most recent calibration files and selected the EPIC events with standard filtering expressions. Time intervals of high background were removed with the \textsc{espfilt} task with standard parameters, thus yielding net exposure times of 7.56  ks (pn), 18.78 ks (MOS1) and 23.41 ks (MOS2).  
We selected single- and multiple-pixel events (\textsc{pattern}$\leq$4 for the pn and $\leq$12 for the MOS cameras); out-of-time events were also removed following the standard procedure\footnote{\url{http://www.cosmos.esa.int/web/xmm-newton/sas-thread-epic-oot}}.

\begin{figure}[ht]
 \centering
 \includegraphics[trim=0cm 0cm 0cm 5cm,clip,width=1\textwidth]{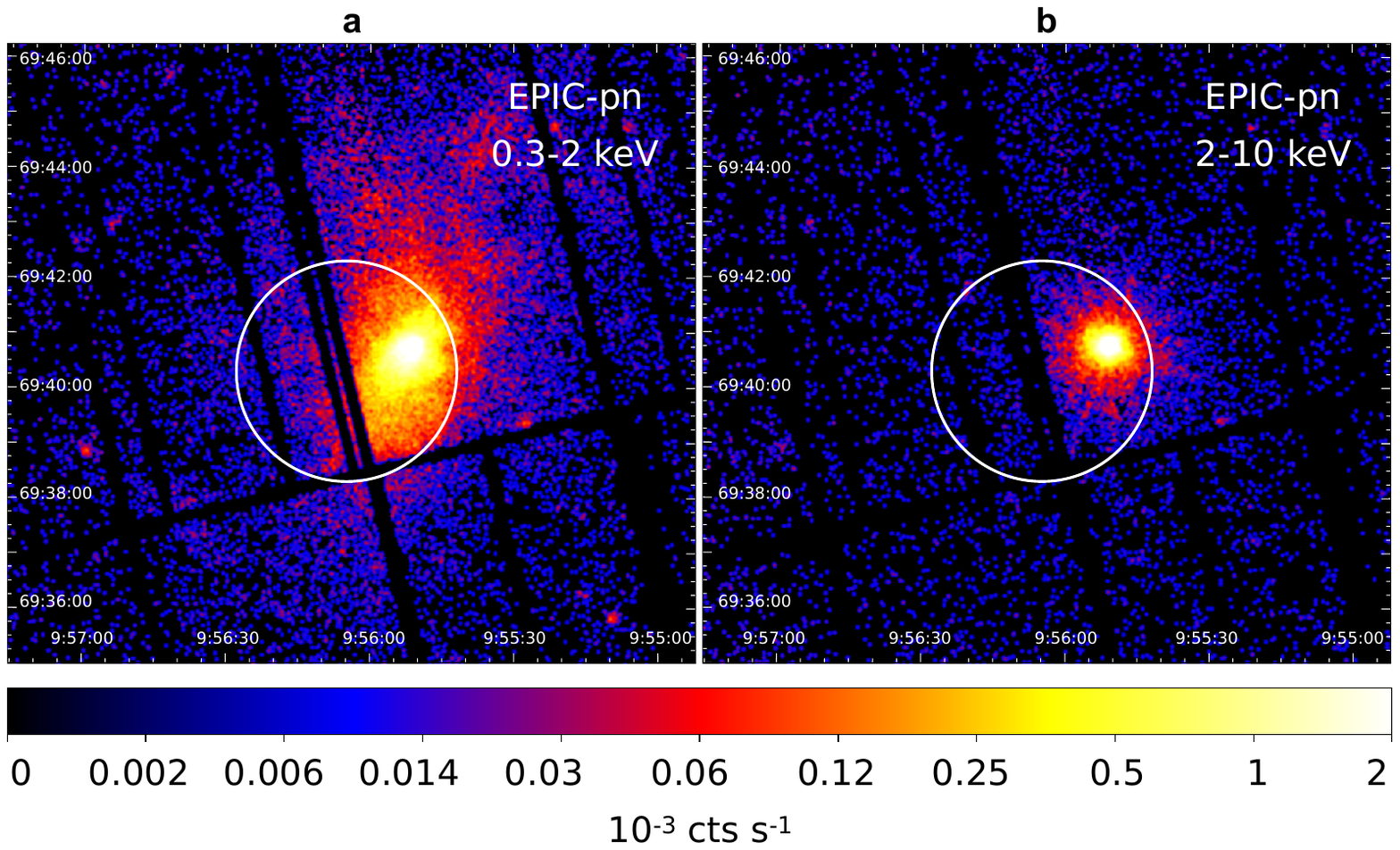}
   \caption{ \textbf{EPIC-pn images of M82}. The   exposure-corrected images refer to the $0.3-2$ keV (\textbf{a}) and $2-10$ keV (\textbf{b}) energy ranges. The 90\% c.l.\ error circle of \grb\ has a radius of 2 arcmin.   }
  \label{fig:images}
\end{figure}

Figure~\ref{fig:images} shows the pn exposure-corrected images in the soft ($0.3-2$ keV) and  hard ($2-10$ keV) energy ranges. 
The images are dominated by the diffuse X-ray emission of the M82 galaxy, especially in the soft X-ray band. 
We compared the  images obtained in this observation with those of all the previous pointings on M82 available in the \xmm archive (Obs.ID 011229, 020608, 056059, 065780, 087094, 089106). We did not find any evidence for the appearance of a new source.
In our observation, the total emission between $2-10$ keV from a circular region of radius 1.5 arcmin  at the center of the galaxy (coordinates \coord{09}{55}{50}{\!.9}{+69}{40}{47}{}; J2000) can be described 
by an absorbed power law with $N_{\rm H}=(1.0\pm0.1)\times 10^{22}$ cm$^{-2}$, photon index $\Gamma=2.01\pm0.04$ and unabsorbed flux $F_{2-10\,\rm keV} = (2.88\pm0.04)\times 10^{-11}$ \flux. The flux measured in previous \xmm observations from the same extraction region and the same spectral model varied between   $\sim\!1.2\times 10^{-11}$ \flux and $\sim\!5.4\times 10^{-11}$ \flux,  indicating the presence of one or more variable sources that cannot be individually resolved with the \xmm spatial resolution.

To compute the upper limit for a new point source  in the \grb\  error box, we applied the \textsc{eupper} task to the $2-10$ keV images. We used a circle of radius $15''$ for the source extraction  and a concentric annulus of radii $22.5''$ and $30''$ for the background. The derived upper limits have been corrected for the fraction of source counts falling outside the extraction circle. 
Figure~\ref{fig:xmm_cr_upplim} shows the 3-$\sigma$ upper limits obtained by applying this procedure on a grid of positions with a step of $5''$.

We used the appropriate EPIC response matrices to convert the count rates of each camera to  fluxes in the $0.3-10$ keV and $2-10$ keV energy ranges, with the assumption of an absorbed power-law spectrum with photon index $\Gamma=2$ and $N_{\rm H}=6.5\times 10^{20}$ cm$^{-2}$ 
Finally we combined the upper limits of the three cameras to obtain  an EPIC upper limit for each energy band. The results are shown in Figure~\ref{fig:xmm_fl_upplim}. About 60\% of the error circle
has a 3-$\sigma$ upper limit of $F_{2-10\,\rm keV}=1.2\times 10^{-14}$ \flux and $F_{0.3-10\,\rm keV}=4.0\times 10^{-14}$ \flux. 
In less than 10\% of the error circle, the upper limit is worse than $7\times 10^{-14}$ \flux ($2-10$ keV) and $1.6\times 10^{-13}$ \flux ($0.3-10$ keV).
At the distance of M82, these fluxes correspond to luminosities of  $1.9\times10^{36}$   and $6.2\times10^{36}$ \lum, respectively.

\begin{figure}[ht]
 \centering
 \includegraphics[trim=0cm 0cm 0cm 5cm,clip,width=1\textwidth]{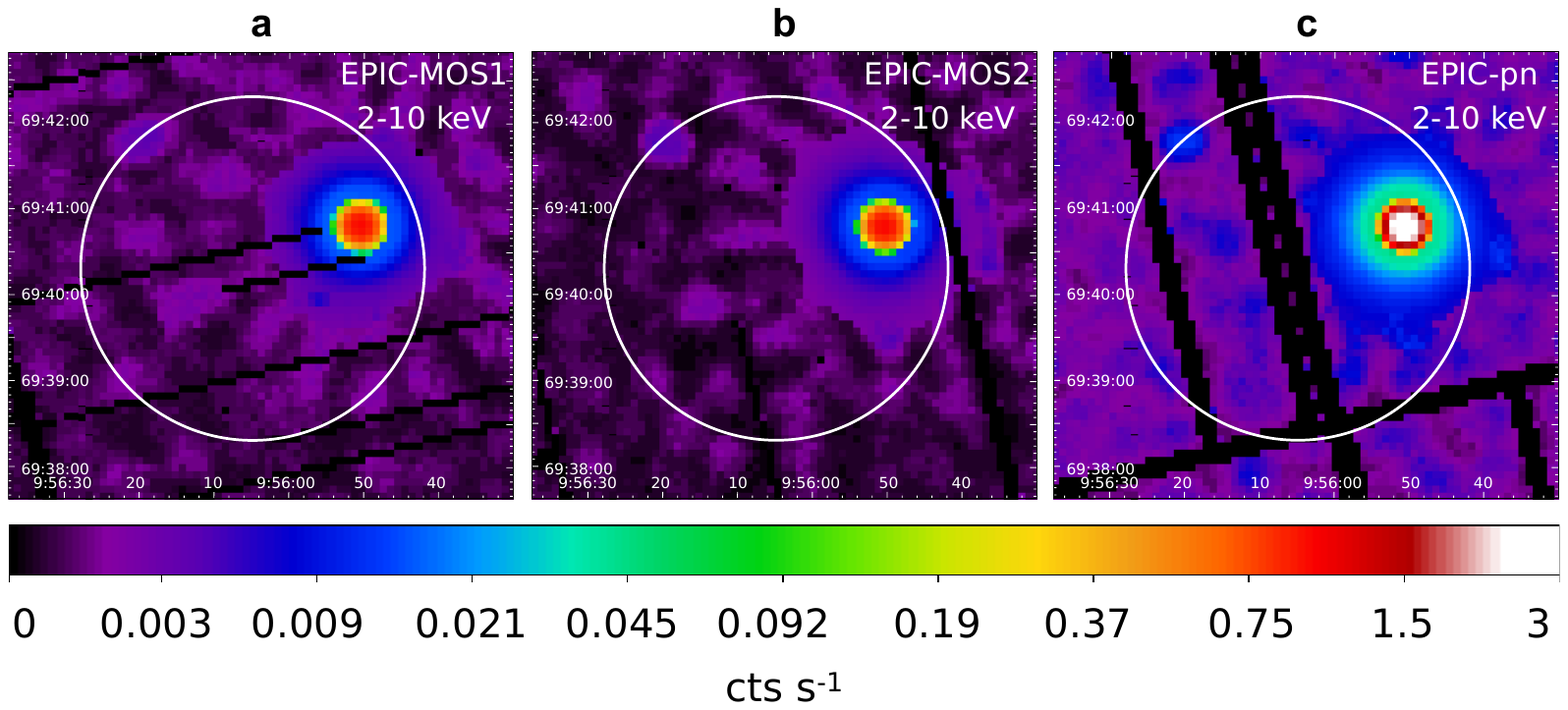}
 \caption{  \textbf{Maps of count rate upper limits.}  The figure gives 3-$\sigma$ upper limits on the $2-10$ keV count rates of the EPIC-MOS1 (\textbf{a}), EPIC-MOS2 (\textbf{b}) and EPIC-pn (\textbf{c}) cameras. The 90\% c.l.\ error circle of \grb\ has a radius of 2 arcmin.   }
  \label{fig:xmm_cr_upplim}
\end{figure}

 \begin{figure}[ht]
 \centering
 \includegraphics[trim=0cm 0cm 0cm 3cm,clip,width=1\textwidth]{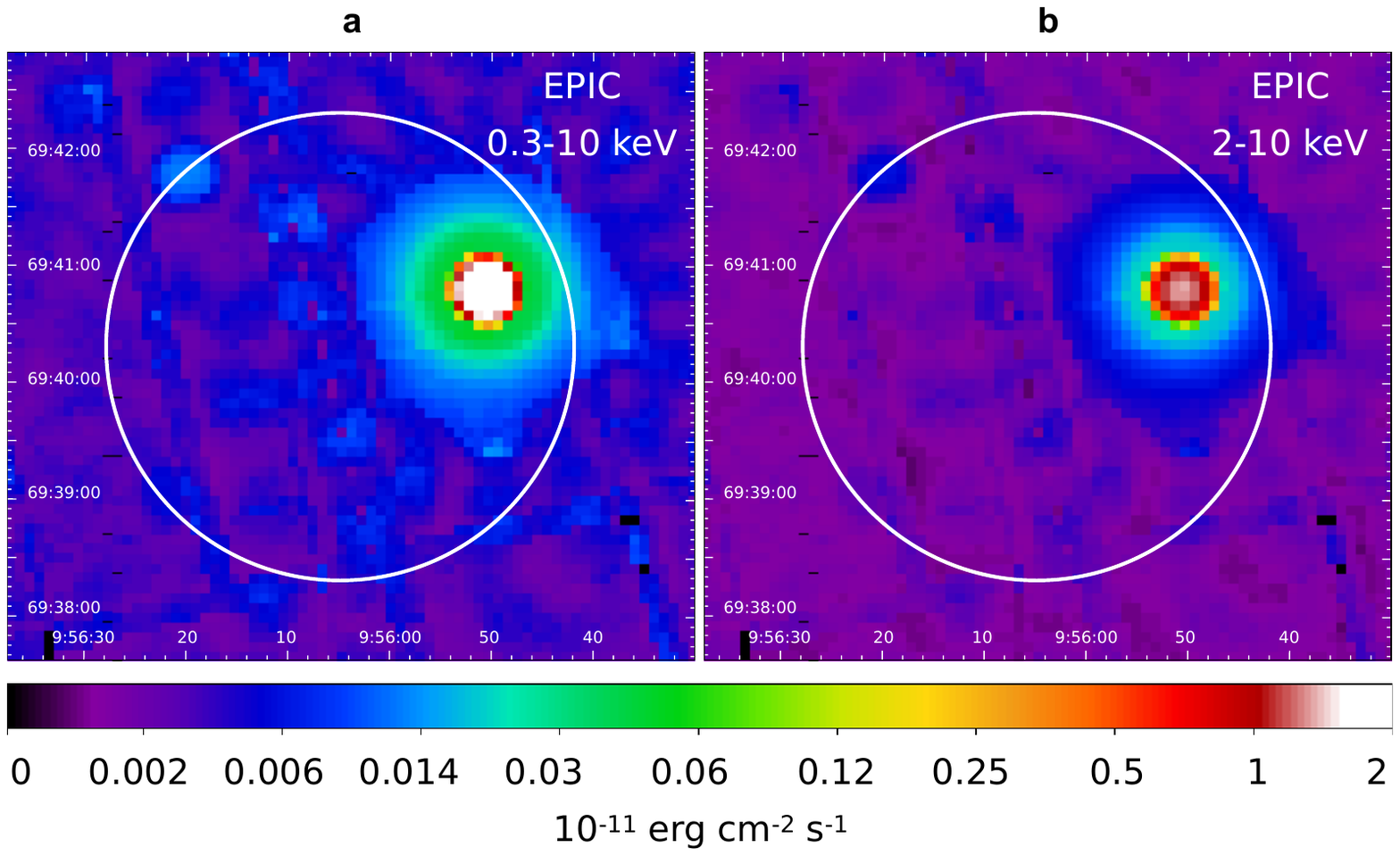}
   \caption{ \textbf{Maps of flux upper limits.}  The figure gives 3-$\sigma$ upper limits on the  fluxes in the $2-10$ keV (\textbf{a}) and $0.3-10$ keV (\textbf{b}) energy range,  obtained by combining the three maps of \ref{fig:S2}. The 90\% c.l.\ error circle of \grb\ has a radius of 2 arcmin.  }
  \label{fig:xmm_fl_upplim}
\end{figure}

\begin{figure}[ht]
 \centering
 \includegraphics[width=0.8\linewidth]{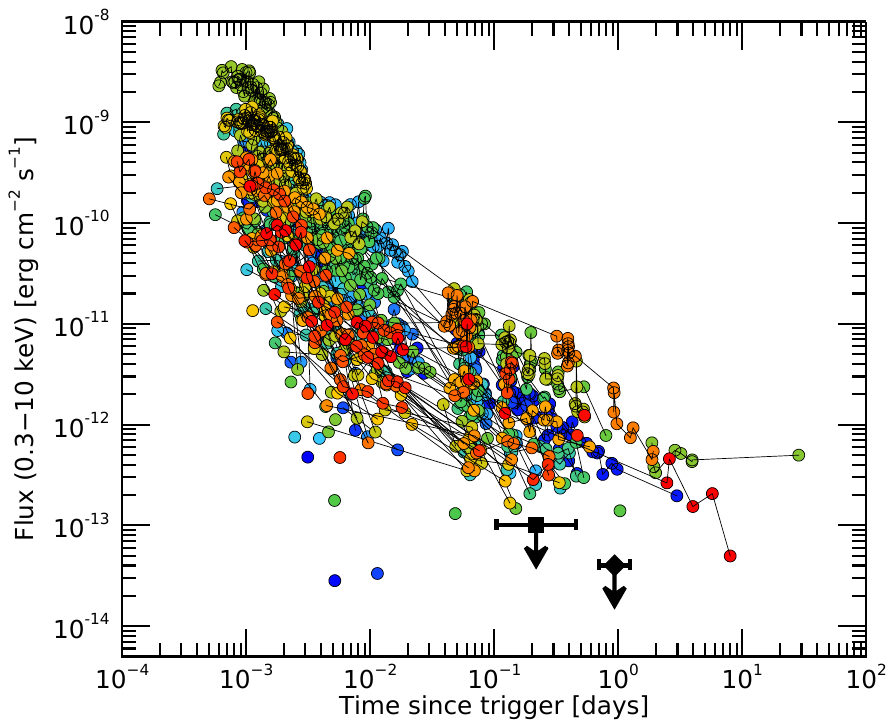}
   \caption{\textbf{X-ray light curves of short GRB afterglows.} The \swift/XRT (black square) and  \xmm/EPIC (black diamond) 3-$\sigma$ upper limits of \grb\ are indicated.  }
  \label{fig:afterglows}
\end{figure}
  
Figure~\ref{fig:afterglows} shows the comparison between the X-ray light curves of short  GRBs (defined such as T$_{90}\leq 2$ s), obtained from the \swift GRB catalog\footnote{\url{http://swift.gsfc.nasa.gov/archive/grb_table}} and the 3-$\sigma$ upper limits we derived from the \swift and \xmm observations of \grb.


\section{Optical observations}
\label{sec:optical}

Multi-filter follow-up optical observations of \grb\ were carried out with the wide-field Schmidt telescopes sited in INAF observatories of Padova (Asiago, Italy), Abruzzo (Campo Imperatore, Italy) and with the 3.6-m Telescopio Nazionale Galileo (TNG, Canary Islands, Spain) between about 5 and 12 hours from the event T$_0$. We also took additional VRI images about 7 hours after T$_0$ with the 120 cm Newton telescope located at the Observatoire de Haute Provence (OHP, France).
The log of the observations is reported in Table~\ref{tab_opt_log}. Image reduction was carried out following the standard procedures: subtraction of an averaged bias frame and division by a normalized flat frame. Astrometry was performed using the Pan-STARRS\footnote{\url{http://outerspace.stsci.edu/display/PANSTARRS/}} catalogue. We carried out image subtraction with respect to SDSS\footnote{\url{http://www.sdss.org}} and telescope archival templates using the \textit{HOTPANTS} (High Order Transform of Psf ANd Template Subtraction code\cite{Becker2015}) package to find and pinpoint variable sources within the \integral error circle.
Aperture and PSF-matched photometry were performed using the DAOPHOT package\cite{Stetson1987} and the {\sc STDPipe}\cite{stdpipe} and the {\sc photutils}\footnote{\url{http://github.com/astropy/photutils}} \cite{photutils} packages for OHP/T120 images.
To minimize any systematic effect, we performed differential photometry with respect to a selection of local isolated and non-saturated reference stars from the Pan-STARRS\footnote{\url{http://outerspace.stsci.edu/display/PANSTARRS/}}. 
The presence of the M82 galaxy makes the background highly non-uniform within the \integral error circle. Magnitude upper limits have been computed for two regions (see Table~\ref{tab_opt_log}): one (most conservative) corresponding to the centre of the M82 galaxy (\coord{09}{55}{53}{\!.75}{+69}{40}{53}{\!.9}; J2000) and another outside the core of the galaxy light (\coord{09}{56}{11}{\!.31}{+69}{39}{09}{\!.6}; J2000).

\begin{table*}
\caption{\textbf{Log of optical observations of \grb.} Magnitudes are in the AB system, not corrected for Galactic extinction. 
Upper limits are given at 3-$\sigma$ confidence level for a source within (outside) the M82 galaxy.
}
\centering          
\begin{tabular}{ccccccc}
\hline
UT observation                &  Exposure           & T - T$_0$  &  Telescope  &  Magnitude	     & Filter \\
(start - stop)                &  (s)                &  (days)    &	      & 		    &	     \\ 
\hline
2023-11-15 20:40:03 - 20:54:15 &  $ 5 \times 180$ s &  0.211     &   Asiago	     & $> 19.1 (> 20.1)$   & $g$    \\
2023-11-15 20:54:59 - 21:12:11 &  $ 5 \times 180$ s &  0.221     &   Asiago	     & $> 19.0 (> 20.1)$   & $r$    \\
2023-11-15 21:12:45 - 21:27:45 &  $ 5 \times 180$ s &  0.234     &   Asiago	     & $> 18.3 (> 19.8)$   & $i$    \\

2023-11-15 21:38:00 - 22:16:30 &  $ 7 \times 300$ s &  0.251     &  Campo Imperatore & $> 14.8 (> 18.9)$   & $z$    \\
2023-11-15 22:38:25 - 21:31:30 &  $ 9 \times 300$ s &  0.293     &   Campo Imperatore & $> 16.5 (> 19.8)$   & $i$    \\
2023-11-15 23:33:32 - 21:38:00 &  $ 7 \times 300$ s &  0.331     &   Campo Imperatore & $> 18.8 (> 20.7)$   & $g$    \\

2023-11-15 22:35:32 - 22:57:23 & $1\times 1400$ s & 0.299 & OHP & $>17.3(>21.6)$ & $r$\\
2023-11-15 23:04:02 - 23:14:02 & $1\times600$ s & 0.314 & OHP & $>18.0(>22.0)$ & $g$\\
2023-11-15 23:14:30 - 23:19:30 & $1 \times 300$ s & 0.320 & OHP & $>17.1(>20.1)$ & $i$\\

2023-11-16 03:25:04 - 03:38:16 &  $ 5 \times 120$ s &  0.492     &  TNG 		     & $> 20.0 (> 24.0)$   & $r$    \\
2023-11-16 03:40:24 - 03:59:09 &  $ 7 \times 120$ s &  0.503     &  TNG 		     & $> 18.9 (> 23.8)$   & $i$    \\
2023-11-16 04:00:28 - 04:19:17 &  $ 7 \times 120$ s &  0.530     &  TNG 		     & $> 18.7 (> 22.7)$   & $z$    \\
\hline
\end{tabular}
\label{tab_opt_log}
\end{table*}

Our   upper limits in the optical bands (Table~\ref{tab_opt_log}) exclude the majority of cosmological short GRB afterglows. If occurred at the distance of M82, in a position located inside (outside) the galaxy, a short GRB would have produced an optical afterglow   more than 12 (16) magnitudes brighter than our limits (see Fig.~\ref{fig:afterglows_opt}). 
These   upper limits are about 9 (13) magnitudes fainter than the expected brightness of a kilonova like AT2017gfo occurring at the distance of  M82  inside (outside) the galaxy. Since kilonovae associated to short GRBs can display significant differences in their luminosity, spectral properties, and temporal evolution\cite{Rossi2020}, we simulated a set of   light curves with the POSSIS code\cite{Bulla2019} using the parameter range  reported in \cite{Ferro2023}. We found that even the faintest event, placed at the distance of M82, is still about 5 (9) magnitudes brighter than our upper limits.

\begin{figure}[ht]
 \centering
 \includegraphics[width=0.8\linewidth]{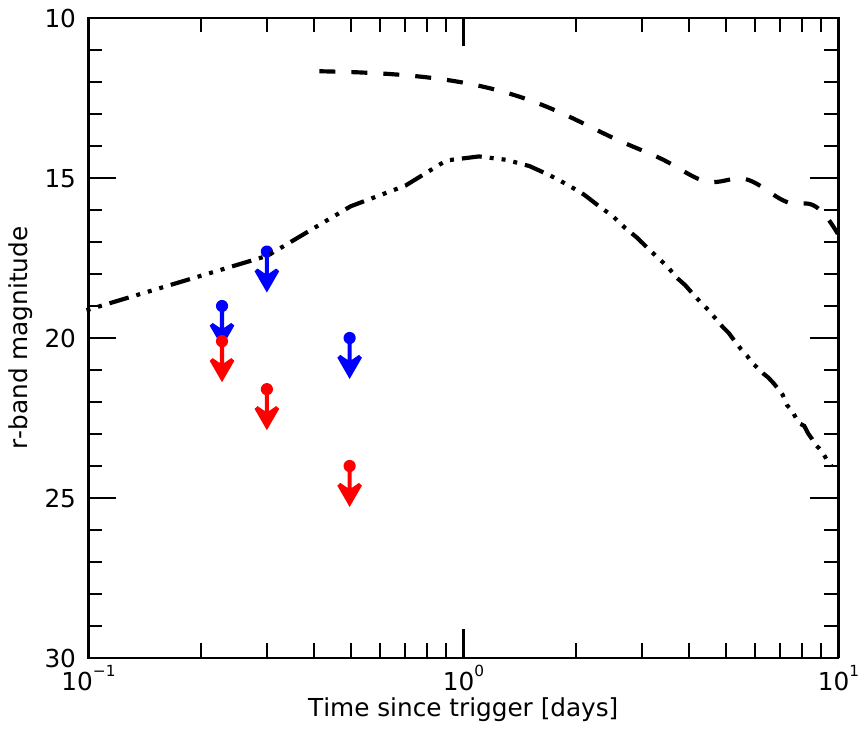}
   \caption{ \textbf{Optical light curves of kilonovae.} The r-band light curves of AT2017gro and of the faintest red kilonova (simulated with the POSSIS code) are shown with dashed and dashed-dotted lines, respectively, assuming the M82 distance (3.6 Mpc). The magnitude 3-$\sigma$ upper limits obtained for a position inside (outside) the M82 galaxy are shown as blue (red) arrows. }
  \label{fig:afterglows_opt}
\end{figure}

\begin{figure}[ht]
 \centering
 \includegraphics[trim=1.0cm 0cm 1.3cm 0cm,clip,width=0.49\linewidth]{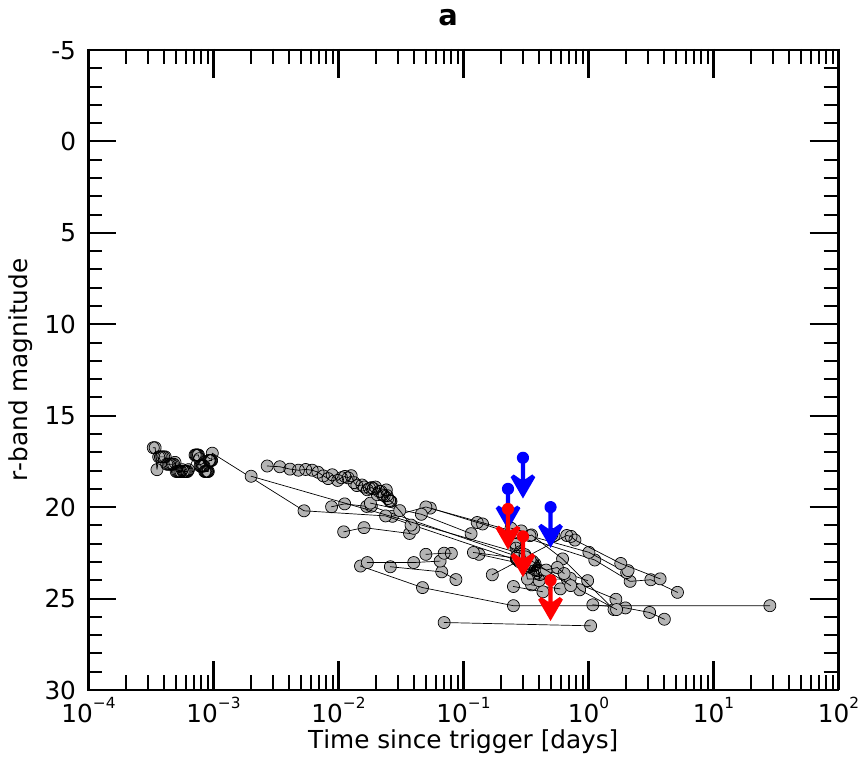}
 \includegraphics[trim=1.0cm 0cm 1.3cm 0cm,clip,width=0.49\linewidth]{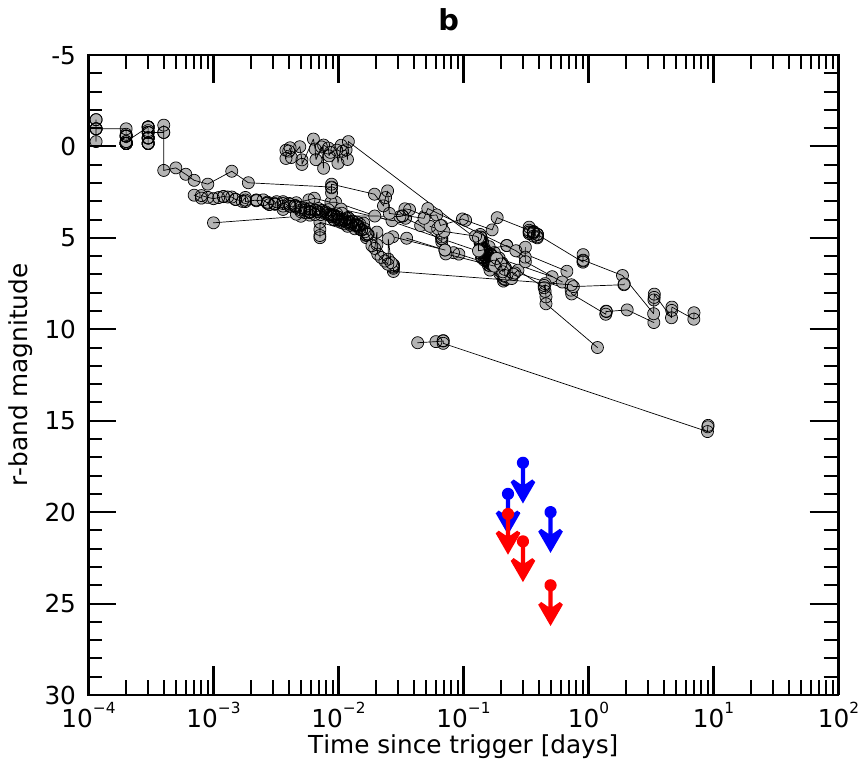}
   \caption{ \textbf{Optical light curves of short GRB afterglows.} The observed light curves are shown in \textbf{a}, while \textbf{b} shows the light curves of those GRBs which have a measure of redshift, rescaled to the M82 distance (3.6 Mpc). The 3-$\sigma$ upper limits obtained for a position inside (outside) the M82 galaxy are shown as blue (red) arrows.  }
  \label{fig:afterglows_opt}
\end{figure}

\section*{References}
\vspace{1cm}

\end{methods}

\begin{addendum}

\item We thank the ESA  Mission Scientists Jan-Uwe Ness and Norbert Schartel for approving and quickly implementing the \integral and \xmm ToO observations. This work is based on observations with \integral and \xmm\ , ESA missions with instruments and science data centres funded by ESA member states, and with the participation of the Russian Federation and the USA. It is also based on observations made with the Italian Telescopio Nazionale Galileo (TNG) operated on the island of La Palma by the Fundación Galileo Galilei of the INAF (Istituto Nazionale di Astrofisica) at the Spanish Observatorio del Roque de los Muchachos of the Instituto de Astrofisica de Canarias.
This paper includes optical data taken with the Schmidt 67/92 telescope operated by INAF Osservatorio Astronomico di Padova (Mt. Ekar, Asiago).
This work received financial support from INAF through the Magnetars Large Program Grant (PI S.Mereghetti) and from the GRAWITA Large Program Grant (PI P. D'Avanzo). 
JCR, AB, SM and PU acknowledge financial support from ASI under contract n. 2019-35-HH.0. 
FO acknowledges support from MIUR, PRIN 2020 (grant 2020KB33TP) ``Multimessenger astronomy in the Einstein Telescope Era (METE).

\item[Authors contributions] 

All authors reviewed the manuscript and contributed to the source interpretation.
SM coordinated the work and the interpretation of the results, contributed to the analysis of the \integral and \xmm data, and wrote most of the manuscript. 
RS and EA contributed to write the main part of the paper.
DP and  JCR  carried out most of the  \integral data analysis.
DG, CF, EB, LD and VS routinely contribute to the operation of the IBAS software and participated to the near real time INTEGRAL analysis.
PDA coordinated the analysis of the optical data from Italian telescopes.
MR  analysed the \xmm data and contributed to the INTEGRAL spectral analysis.
SC analysed the \swift data.
MT contributed to the software for the burst search in archival data.
DT, WT  and CA coordinated the observation and the analysis of the optical data taken at OHP.
LT analysed  the optical data taken with the Schmidt 67/92 telescope in Asiago under the Large Program  "Search and characterisation of optical counterparts of GW triggers”  (P.I. Tomasella). AR and EC triggered, reduced and analysed the observations at the Asiago Schmidt telescope. 
RB and MF provided the short GRB afterglows and kilonovae observed and simulated optical light curves. 

\item[Competing Interests] The authors declare no competing interests.

\item[Correspondence] Correspondence and requests for materials should be addressed to Sandro Mereghetti (email: sandro.mereghetti@inaf.it).

\item[Data Availability]  The data of the INTEGRAL, \xmm and \swift satellites are publicly available in the respective online archives \\
(https://www.isdc.unige.ch/integral/archive,\\
 https://www.cosmos.esa.int/web/xmm-newton/xsa,\\
https://swift.gsfc.nasa.gov/archive/ )\\.
Optical data are available upon reasonable request.
 
\item[Code Availability] The software used for the data analysis is public and can be retrieved at \\
 https://www.cosmos.esa.int/web/xmm-newton/sas,\\
 https://www.isdc.unige.ch/integral/analysis Software, \\ 
 https://heasarc.gsfc.nasa.gov/xanadu/xspec/

\end{addendum}

\end{document}